\documentclass[pra,aps,superscriptaddress,twocolumn]{revtex4}

\usepackage{graphicx}
\usepackage{wrapfig}
\usepackage{amsfonts}
\usepackage{amssymb}
\usepackage{amsmath}
\usepackage{bm}
\usepackage{textgreek}
\usepackage{color}
\newcommand{\Fig}[1]{Fig.\,\ref{#1}}
\newcommand{\Sec}[1]{Sec.\,\ref{#1}}

\newcommand{\Ab}{A\textbeta}
\newcommand{\Comp}{\ensuremath{{\cal C}}}
\newcommand{\rcm}{cm$^{-1}$}

\begin{document}
\title{Quantitative chemical imaging of amyloid-\textbeta~plaques with Raman micro-spectroscopy in human Alzheimer's diseased brains}
\author{E.\,G.~Lobanova}\email{Moreva@physics.msu.ru}\affiliation{School of Biosciences, Cardiff University, Cardiff CF10 3AX, United Kingdom}
\author{S.\,V.~Lobanov}\affiliation{School of Physics and Astronomy, Cardiff University, Cardiff CF24 3AA, United Kingdom}
\author{K.~Triantafilou}\affiliation{School of Medicine, Cardiff University, Cardiff CF14 4XN, United Kingdom}
\author{W.~Langbein}\affiliation{School of Physics and Astronomy, Cardiff University, Cardiff CF24 3AA, United Kingdom}
\author{P.~Borri}\email{BorriP@cardiff.ac.uk}\affiliation{School of Biosciences, Cardiff University, Cardiff CF10 3AX, United Kingdom}
\begin{abstract}

Alzheimer's disease\,(AD) is a neurodegenerative disorder and the most common cause of dementia in the elderly. The extracellular accumulation of amyloid-\textbeta\,(\Ab) in senile plaques is a principal event in the pathogenesis and there is growing evidence that the dysregulation of lipid pathways is implicated in the disease, however the link between these two is still under study. In this work, we investigated human brain samples, from 11 AD patients and a control cohort of age-matched subjects without AD, using label-free chemically-specific Raman micro-spectroscopy. The collected image data were quantitatively analysed using an efficient quantitative unsupervised/partially supervised non-negative matrix factorization method, to retrieve the concentration maps and spectra of the samples' chemical constituents. Significant changes in lipid composition as well as increased concentrations of oxidative stress bio-markers were observed in AD tissues compared to the control. In particular, the analysis revealed accumulations of two different lipid components throughout the entire \Ab~plaque (highest at the fibrillar core and lowest at the rim). One component is attributed to cholesteryl esters with saturated long-chain fatty acids\,(FAs), found to co-localise with aggregated \Ab~peptides in multi-clusters of about 3\,$\mu$m average size, which are in turn bundled to each other forming one macro-aggregate templating the \Ab~plaque. The other component is attributed to a mixture of fibrin protein and arachidic acid, which are a
hallmark of inflammation, and was found to be abundant in fibrillar and core-only~plaques. This fibrin/arachidic acid component showed multiple domains of different sizes, ranging from 1 to 5\,$\mu$m, linked to each other and forming a large structure of shape similar to the plaque. Lipid aggregates and \Ab~plaque cores for the AD group were also found to be co-localised with significant accumulations of $\beta$-carotene and magnetite~(Fe$_3$O$_4$), indicating a high level of oxidative damage implicated in AD human brains.  In addition, a non-\Ab~protein component was observed to co-deposit with the fibrillar core of hippocampal \Ab~plaques, and was attributed to a collagen-like amyloidogenic component~(CLAC). Overall, our results reveal abnormal deposits of saturated cholesteryl esters, \Ab~fibrils, arachidic acid, fibrin, CLAC,
$\beta$-carotene and magnetite, co-localising in \Ab~plaques of AD human brains. This finding opens perspectives for new anti-inflammatory and antioxidant drug strategies, designed to restore brain homeostasis as potential therapeutics of AD. We also demonstrate by the means of spatial concentration histograms that these identified species separate AD from non-demented control brains, beneficial for AD diagnosis.

\end{abstract}
%
%
\date{\today}
\maketitle

\section{Introduction}\label{sec:intro}

Alzheimer's disease\,(AD) is an age-related neurodegenerative disorder,
characterized by the extracellular accumulation of \Ab~peptide in senile plaques
and the intracellular deposits of the microtubule-associated protein tau in the
form of neurofibrillary tangles. According to the 2017 Alzheimer's Association
report, AD affects one in nine Americans over the age of 65 and becomes
increasingly prevalent with age\,\cite{AlzheimersAssociation2017}. Alongside
obesity and diabetes, AD is a modern disease with consistently increasing number
of patients. Despite being considered as a 21st century epidemic, there is still
no cure for AD and the etiology of the disease remains obscure. A number of
hypotheses such as amyloid cascade, lipid dysregulation, oxidative stress, and
inflammation have been proposed for the AD pathogenesis, suggesting that AD is a
multifactorial complex disorder.

\Ab~plaques, lipids and neuroinflammation can be considered as the three main
hallmarks of AD. Indeed, it has been shown that the aggregation of \Ab~is
mediated by the interaction with different classes of lipids, including
cholesterol\,\cite{Sparks1994,Refolo2000,Paolo2011},
ganglioside\,\cite{Choo-smith1997,Matsuzaki1999,Duncan1995},
phospholipids\,\cite{McLaurin1997,Lee2002} and fatty acids (FAs)\,\cite{Prasad1998}, which
activates the innate immune system and then induces a cascade of pathological
events such as neuroinflammation and oxidative stress. Interestingly, in highly
oxidative environment, characterized by high level of reactive oxygen species (ROS) as well as redox metal
ions, lipids become oxidized\,\cite{Markesbery1997}. The products of lipid
peroxidation are very reactive with other biomolecules and, consequently,
neurotoxic\,\cite{Esterbauer1991,Butterfield2011}. As an evidence, it has been
shown that 4-hydroxy-2-trans-nonena\,(HNE), the product of lipid peroxidation,
can covalently modify the histidine side chains of \Ab, leading to increased
aggregation of this peptide\,\cite{Liu2008}. In addition, neprilysin~(NEP), a
major protease that cleaves \Ab~\textit{in vivo}, has also been reported to be
HNE modified in AD brain\,\cite{Wang2003}. On the other hand, there is a line of
evidence that lipids themselves have detergent properties, promoting the
disintegration of amyloid fibrils into toxic oligomers\,\cite{Widenbrant2006,Martins2008}. These free oligomers can bind to
synaptic contacts\,\cite{Deshpande2006}, where they disrupt membrane
homeostasis\,\cite{Milanesi2012}. Altogether, these findings suggest that the
amyloid-lipid complex in \Ab~plaques produces toxic species, which are involved
in AD's pathology.

Despite the importance of this research, the direct molecular analysis of
\Ab~plaques and their lipid inclusions in human AD brains, with quantitative
information on their spatial distribution and chemical composition, is still very
limited. This is partly due to the lack of reliable methods to study lipids with
high spatial resolution and chemical specificity in their native
microenvironments. Of the imaging tools available, fluorescent microscopy\,(FM)
revealed abnormal deposits of cholesterol in cores of mature
\Ab~plaques\,\cite{Mori2001}. However, this technique is based on tagging the
biomolecules of interest, not allowing to study unknown chemical species.
Furthermore, labelling can affect the distribution/composition of the fragile
lipids/amyloid co-arrangements. Among spectroscopy techniques, mass
spectrometry\,(MS) provides chemical information on a wide range of molecular
contents with high specificity, but requires sample ionization as well as
chemical extraction of lipids from tissue samples. For example, liquid
chromatography-MS\,(LC-MS) showed an elevated level of diacylglycerol~($\sim$80$\%\uparrow$),
glucosylceramide~($\sim$43$\%\uparrow$) and galactosylceramide~($\sim$32$\%\uparrow$), as well as decreased phosphatidylethanolamine~($\sim$25$\%\downarrow$) in the prefrontal cortex human brain with AD compared to controls. Furthermore, the AD entorhinal cortex subjects were found to be enriched in lysobisphosphatidic acid~($\sim$80$\%\uparrow$),
sphingomyelin~($\sim$20$\%\uparrow$), ganglioside GM3~(ceramide-N-tetrose-N-acetylneuraminic acid)~($\sim$64$\%\uparrow$), and cholesterol esters\,(CE)~($\sim$70$\%\uparrow$) in the form of CE-16:1, CE-16:0, and CE-18:1, which are typically associated with lipid
droplets\,\cite{Chan2012}. It is possible to combine MS with imaging and achieve
spatial resolution from tens of micron (with MALDI) to sub-micron (with second
ionization mass spectrometry-SIMS). These MS-based imaging modalities have been
applied to image lipid, proteins and metals in
brains\,\cite{Grasso2011,Braidy2014}, however they require sample processing
procedures which are damaging and destructive and can not be applied to living
specimens.

Vibrational micro-spectroscopies techniques such as infrared
absorption\,\cite{Liao2013,Benseny-cases2014} and Raman scattering microscopy
detect the intrinsic vibrations of chemical bonds in molecules, and thereby
provide a non-invasive, label-free chemically specific imaging method. Infrared
microscopy has limited spatial resolution due to the long wavelength of the
infrared light. Conversely, Raman micro-spectroscopy can offer sub-micron
spatial resolution by using light in the visible range. Notably, in its
non-linear version, coherent Raman scattering\,(CRS) microscopy provides
additional improvements in 3D spatial resolution and acquisition speed compared
to spontaneous Raman, and has been recently applied to study brain
tissues\,\cite{Lee2015,Freudiger2008,Kiskis2015}. However CRS suffers from
limited signal-to-noise ratio in the vibrational fingerprint range
(700-1800\,\rcm), and in turn reported studies of CRS microscopy of brain
tissues examined only the CH-stretch vibrational range (2700-3100\,\rcm),
which is notoriously more congested than the fingerprint region and thus less
chemically specific. Surprisingly little work has been reported to date on the
chemical profile of AD brains by confocal Raman micro-spectroscopy
\,\cite{Michael2014,Michael2017}, and a comprehensive quantitative analysis of the
spatially-resolved chemical composition of the lipid-amyloid complex, covering
the whole spectral range from fingerprint to CH-stretch, in unstained AD human
brain tissues is still missing.

In this work, we report an extensive study of the spatial distribution and
chemical composition of \Ab~plaques, measured by label-free spontaneous Raman
micro-spectroscopy in fixed human brain slices from AD patients and age-matched controls and quantified by hyperspectral image factorization method.

\section{Experimental Section}
\subsection{Tissue preparation}\label{Sec:Tissue}
\begin{table*}\caption{Clinical characteristics of study participants}
\begin{tabular}{| c | c | c | c | c | c |}
\hline
\: Case Number \: &	 \: Age in years  \: & \:	Gender \: & \:	Neuropathological diagnosis \: & \:	The degree of AD pathology~\cite{Hyman2013}  \: & \:	Brain region \: \\
\hline
Ah1 &	92 &	F &	AD &	Severe &	Hippocampus \\
Ah2 &	87 &	F &	AD &	Severe &	Hippocampus \\
Ah3 &	73 &	M &	AD &	Modereate &	Hippocampus \\
Ah4 &	70 &	M &	AD &	Mild/Moderate &	Hippocampus \\
Ah5 &	85 &	F &	AD &	Moderate &	Hippocampus \\
Ac1 &	75 &	M &	AD &	Moderate &	Frontal Lobe \\
Ac2 &	82 &	F &	AD &	Severe &	Frontal Lobe \\
Ac3 &	85 &	F &	AD &	Severe &	Frontal Lobe \\
Ac4 &	80 &	M &	AD &	Moderate &	Frontal Lobe \\
Ac5 &	76 &	M &	AD &	Moderate &	Frontal Lobe \\
Ac6 &	77 &	M &	AD &	Moderate &	Frontal Lobe \\
Ch1 &	59 &	M &	Control &	Normal &	Hippocampus \\
Ch2 &	61 &	F &	Control &	Normal &	Hippocampus \\
Cc1 &	64 &	M &	Control &	Normal & 	Frontal Lobe \\
Cc2 &	60 &	F &	Control &	Normal &	Frontal Lobe \\ \hline
\end{tabular}
\end{table*}
{Human brain tissue was obtained from the Thomas Willis Oxford Brain Collection~(TWOBC), which is part of the Medical Research Council~(MRC) Brain Bank Network, with the ethical approval of the Research Tissue Bank provided by the Oxford Brain Bank (OBB) Access Committee.
Brain tissue samples were recruited from 11 individuals, who at the
time of death had AD, and from 4 age-matched humans, which were not
diagnosed with AD based on histological examination~(see Table~1).
The AD brain cohort met the standard criteria for AD's neuropathologic diagnosis.
A total of 5 hippocampal and 6 cerebral cortex diseased samples were
investigated, representing the hippocampus of the first five AD
subjects and the frontal lobe of the other six ones. As control, 2
hippocampal and 2 cerebral cortex samples from the corresponding
brain regions of non-demented subjects were included in this study.
Brain samples were fixed with formalin, paraffin embedded, sectioned
into 5~$\mu$m thick slices and then mounted on standard glass
slides. For Raman micro-spectroscopy measurements,
formalin-fixed-paraffin-embedded brain sections were deparaffinized
using a standard dewaxing protocol: removing the paraffin in xylene (two cycles for 20~minutes each), and rehydrating in a sequence of 100\%, 90\% and 70$\%$ ethanol
for 5~minutes each. The slices were then covered with phosphate buffered
saline\,(PBS) and a standard glass coverslip was attached on top. 
Finally, the mounted samples were sealed with nail varnish applied around the edges of the coverslip. After Raman measurements, the samples were reopened and stained
with Thioflavin-S for amyloid fibrils to confirm plaque
locations. In the following, we refer to AD and
control (C) samples, numbered by individual, as Ah1,...,\,Ah5 and
Ch1,\,Ch2, respectively, studied in the hippocampus; and as Ac1,...,\,Ac6, and Cc1,\,Cc2, studied in the cerebral cortex. 
A total of 61~\Ab~plaques were examined: 30 in the five hippocampal brain samples and 31 in the six cerebral cortex
samples from the 11 AD subjects. Specifically, there were: 10, 9, 5, 1, and 5 plaques in the Ah1, Ah2, Ah3, Ah4, and Ah5 hippocampal samples, respectively; and 3, 13, 7, 2, 3, and 3 plaques in the Ac1, Ac2, Ac3, Ac4, Ac5, and Ac6 cortical samples.

\subsection{Optical micro-spectroscopy}\label{Sec:Microspec}

Raman imaging was carried out using a home-built multimodal
laser-scanning microscope based on an inverted Nikon Ti-U stand,
coupled to a Horiba Jobin-Yvon iHR 550 imaging spectrometer (300
grooves/mm grating) and an Andor CCD Newton DU-971N-BV detector
(1600x400\,pixels, 16\,$\mu$m pixel size). A laser source (Laser Quantum GEM 532\,nm) was
used for excitation. The 532\,nm laser excitation was filtered by a Semrock
LL01-532 clean-up filter, and coupled into the microscope by a dichroic mirror (Semrock LPD01-532RS). Raman scattering
was collected in epi-direction, filtered with a long pass filter (Semrock BLP01-532R), and dispersed
by the imaging spectrometer. The spectral
resolution of the system was 5\,\rcm\ with 30~$\mu$m width of the
spectrometer entrance slit, and the measured Raman spectral range covered 330\,\rcm\ to 4050\,\rcm.

Raman imaging for all brain samples was achieved by raster-scanning the sample, using a
motorized stage (Prior ProScan III XYZ Controller), through the focal point of
the laser beam over a region of interest of about $50\times50\,\mu$m$^2$ with a step size of
1\,$\mu$m resulting in a 50$\times$50$\times$1600 points hyperspectral Raman
image with a 1 second exposure time per spectrum. All images were
taken with 100~mW laser power at the sample, using a 20$\times$\,0.75 NA dry objective (Nikon MRD00205) and 1$\times$\,tube lens. Confocal Raman detection was provided by a horizontal slit (Thorlabs VA100), and the vertical input slit of the spectrometer, both located in an intermediate image plane.
The estimated spatial resolution was 0.45\,$\mu$m in-plane and 0.6\,$\mu$m axially. 

Prior to Raman measurements, an overview image for each sample across $10.4 \times 6$\,mm$^2$, made of $24 \times 18$ tiles with $0.43 \times 0.33\,$mm$^2$ tile size, was acquired using differential interference contrast\,(DIC). In the DIC, a de-Senarmont compensator provided a phase offset of 20 degrees and a 0.72 NA dry condenser (Nikon MEL56100) with a DIC module (Nikon MEH52400) was used for illumination, combined with a matched DIC slider (Nikon MBH76220) in the objective, giving a shear of about 0.24\,$\mu$m.
A Hamamatsu ORCA 285 camera ($1344 \times 1024$ pixels of 6.45\,$\mu$m size) at 100 ms frame exposure time was used for detection at lowest gain (4.45 photoelectrons/count).

For imaging of the Thioflavin-S stained samples, wide-field epi-fluorescence microscopy was implemented in the same microscope set-up using a Prior Lumen 200 (standard) 200W fluorescence illuminator light source, and a filter set consisting of a single band exciter~(Semrock FF01-370/36) transmitting 352-388\,nm, a single band dichroic~(Semrock FF409-Di03) at 409\,nm, and a single band emitter ~(Semrock FF02-447/60) transmitting 417-477\,nm.  Each image was acquired with 200\,ms exposure time and 10\% lamp power.

\subsection{Data analysis and processing}\label{Sec:Data}
Raman hyperspectral data were analysed using quantitative unsupervised/partially supervised unmixing method\,\cite{Lobanova2018b}, consisting of the following basic steps (see also the theory, algorithms and data analysis steps in\,\cite{Lobanova2018b}). 
Firstly, the data were noise-filtered using singular-value decomposition\,(SVD) with automatic divisive correlation, enabling to remove components which are identified as noise using the auto-correlation coefficients for spatial and spectral singular vectors at one pixel shift.
Components with a mean auto-correlation coefficient less than 50\% were attributed to noise and discarded. This cut-off value was suited to differentiate the meaningful components from noise-dominated ones, as verified by visual inspection and discussed in \citep{Lobanova2018b}.
Secondly, broad features in the spectra, which represent fluorescent background, were subtracted using the bottom Gaussian fitting with a standard deviation\,(STD) of 900\,\rcm.
Finally, the efficient quantitative unsupervised/partially supervised non-negative matrix factorization~(Q-US/PS-NMF) method via a fast combinatorial alternating non-negativity-constrained least squares~(ANLS) algorithm was applied to the noise-filtered background-subtracted data. This algorithm \citep{Lobanova2018b} factorizes the spectrally and spatially resolved data into a sum of direct products of non-negative spatial concentration maps and corresponding non-negative spectra, representing  biochemical components of sample composition, showing an exponential convergence towards a global minimum. In the factorizations shown here, the remaining factorization error was about 2\%. In order to compare the concentrations of different chemical species between samples, the component spectra were normalized to have equal integrals, having a value chosen to result in a sum of the spatial concentrations with a mean of one.

In order to separate the contribution of wax residues to the spectral basis, the spectrum of the paraffin-wax compound, measured on one of the paraffin-embedded sample, was used as a fixed spectral component in the Q-US/PS-NMF. The Raman spectrum of wax is also shown in Figs.\,S11 and S33 of the Supporting Information, together with the results of factorization analysis of hippocampal and cortical brain tissues.


\section{Results and Discussion}

Label-free confocal Raman micro-spectroscopy together with hyperspectral image factorization method (see \Sec{Sec:Microspec} and \Sec{Sec:Data}) was used to investigate the biochemical composition
of \Ab~plaques and their environment in AD human brain tissues
from the hippocampus and frontal lobe. The quantitative data
analysis was performed {\it simultaneously} on all data of a given type of brain region (i.e. hippocampus or cortex),
enabling a direct comparison of concentration and spatial distribution of
chemical components in diseased and control samples.

To identify \Ab~plaques, the samples were stained with Thioflavin-S (the
"gold standard" for identifying amyloid fibrils\,\cite{Kelenyi1967}). This was done after Raman imaging, to avoid labelling artefacts and fluorescence background. To identify potential areas containing plaques on label-free AD samples, we used large scale DIC images. DIC contrast allowed us to identify \Ab~plaques regions, which were used for the area selection for Raman measurements of label-free AD samples. To confirm they are plaque-positive, we performed FM of the measured regions on the same samples stained with Thioflavin-S dye.

The factorization analysis revealed that
all hyperspectral images were optimally represented using 20 separate components. Detailed discussion on the components' number selection can be found in\,\cite{Lobanova2018b}.  In brief, when using 19 components,
the algorithm resulted in a higher concentration error, whereas for 21 components one factorization component with a meaningful spectrum which could be attributed to proteins split into two less meaningful ones.
Five of these components were found to be spatially co-localized with the \Ab~plaques, and exhibited spectra resembling those of known chemical species, as
detailed in the following section. We call these components $\Comp_1..\Comp_5$ in the following, sorted in the order of decreasing mean
concentration. Notably, four of them were similar in the hippocampal and cortical brain
samples, with the remaining one was seen only in the
hippocampus.
The other 15 components were not significantly colocalized with \Ab~plaques. Their description and attribution is given in the Supporting Information.

\subsection{Raman spectra of chemical components colocalized with hippocampal \Ab~plaques}\label{sec:RamanAssignment}

The spectra of the 5 chemical components $\Comp_1...\Comp_5$
co-localized with \Ab~plaques in hippocampal samples in the fingerprint
(736-1780\,\rcm) and CH-stretch (2793-3122\,\rcm) regions are given in \Fig{Fig1}. They were obtained analysing data from 30 regions containing \Ab~plaques from AD subjects and 10 regions from the control group (see \Sec{Sec:Tissue}).
The ranges below 736\,\rcm and above 3200\,\rcm are dominated by Raman scattering from the glass slide/coverslip and water, respectively, and were therefore excluded from the analysis. 

Raman spectra of analytical standards, which resemble the component spectra, are shown for comparison. They were identified using Bio-Rad's KnowItAll Vibrational Spectroscopy software with the Raman Spectral Libraries. The Raman spectra of the identified standards shown in this work were then taken from published papers and are cited accordingly in the following.

\begin{figure}
\includegraphics[width=0.95\linewidth]{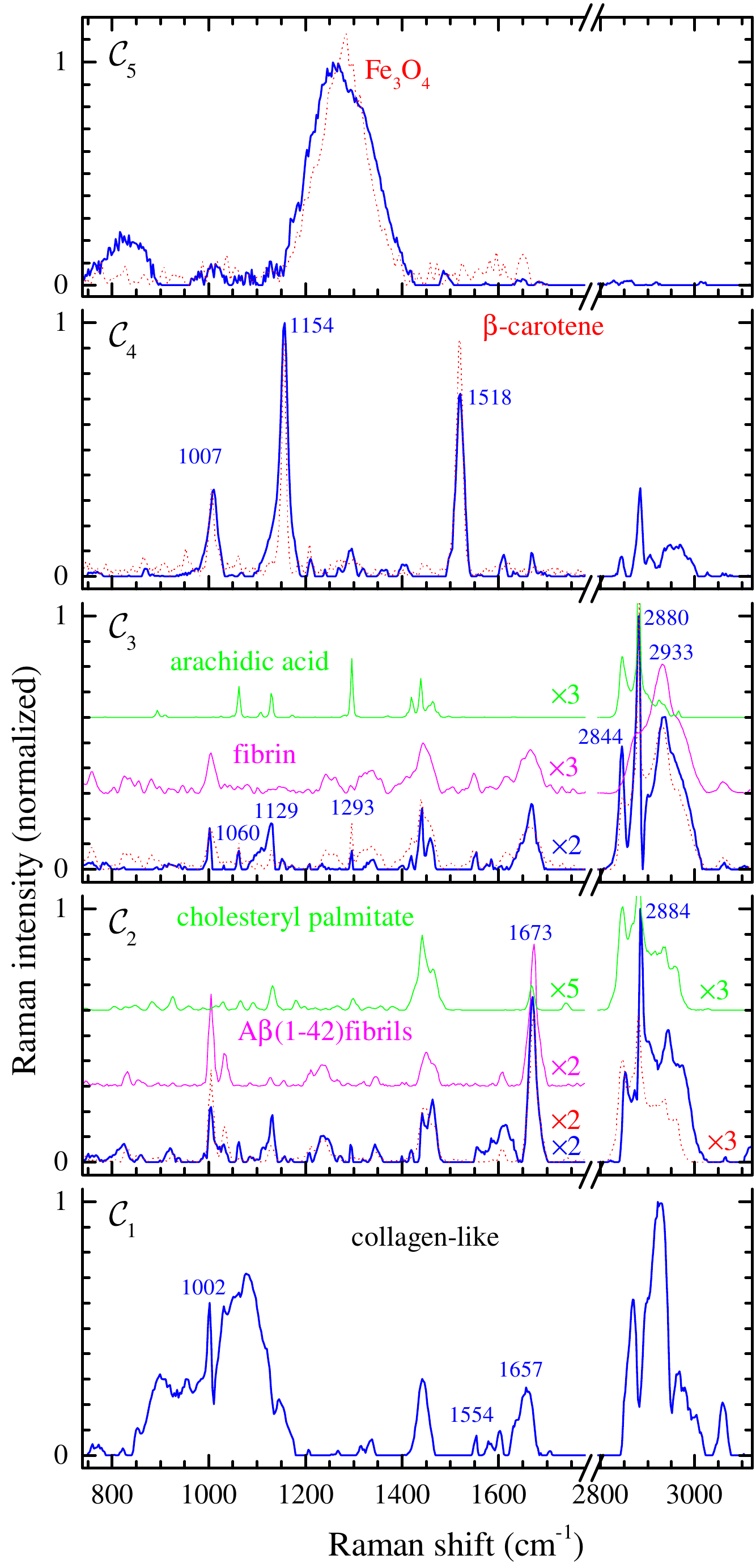}
\caption{Raman spectra (blue solid lines) of the five chemical components $\Comp_1...\Comp_5$ co-localized with hippocampal \Ab~plaques, as indicated. For each component, spectra of mixtures of analytical standards are shown (red dotted lines), with spectra of the constituents shown as green and magenta lines, offset and scaled for clarity.} 
\label{Fig1}
\end{figure}

The spectrum of $\Comp_1$ exhibits a series of characteristic
bands, at 1657\,\rcm~(Amide~I, $\alpha$-helix),
1554~\rcm\,(Amide~II), 1338\,\rcm~(CH$_2$ wagging vibrations
from glycine~(Gly) backbone and proline~(Pro) side chain),
1267\,\rcm~(Amide~III), 1206\,\rcm~(hydroxyproline~(Hyp) and
tyrosine~(Tyr)), 1031\,\rcm~(Pro),
1002\,\rcm~(phenylalanine), 934\,\rcm~(C-C backbone),
852\,\rcm~(Pro, Hyp, Tyr), which are consistent with the Raman
bands of collagen\,\cite{Movasaghi2007}. Furthermore, the spectrum
shows a notable band at 1705\,\rcm, which is characteristic for
the C=O stretching vibration of amino acids. This band might be
originated from Gly-Pro-X and Gly-X-Hyp sequence (where X is any
amino acid), which are the most common motifs of collagen. 

It has been previously reported that senile plaques extracted from AD
human brains contain a collagen-like amyloidogenic component\,(CLAC),
which is associated with collagen~XXV\,\cite{Hashimoto2002}.
\textit{In vitro} studies show that addition of a collagen-like
triple-helical peptide to the amyloidogenic peptide significantly
delays its nucleation and fibril growth\,\cite{Parmar2012}. Relating
to the C=O vibration observed for $\Comp_1$, these studies
hypothesised that the C=O moieties in the repeating Gly-Pro-Hyp
sequences within the triple-helix collagen can bind to backbone
amides or to glutamine/asparagine side chains of the amyloid-like
peptide, hampering its nucleation. Other relationships between
collagen and amyloid suggest the neuroprotective potential of
collagen~VI against \Ab~neurotoxicity by blocking the
association of \Ab~oligomers with neurons, protecting against
fragmentation of \Ab~plaques, thereby preventing
neurotoxicity \cite{Cheng2010}.

The spectra of $\Comp_2$ and $\Comp_3$ show bands characteristic of a
mixture of lipids, with $\Comp_2$ additionally showing bands that can be related to proteins. In particular, the fingerprint region of $\Comp_3$ exhibits bands  at 1060, 1129, and 1293\,\rcm, assigned to C-C skeletal stretching vibrations, and the CH$_2$ twisting mode of fatty acids, respectively\,\cite{Krafft2005,Czamara2015}. The presence of a
strong 2880\,\rcm\ band in the CH-stretch region (the asymmetric CH$_2$ stretch enhanced by the Fermi resonance interaction with the overtones of CH$_2$ and CH$_3$ deformations) combined with a relatively weak band around 2933\,\rcm\ (a combination of CH$_3$ and CH$_2$ asymmetric vibrations enhanced by the broadening and shift of the CH deformations in the liquid phase)\,\cite{DiNapoli2014}, is typical of a saturated lipid in the solid phase.  We find that the spectrum of $\Comp_3$ can
be well described (using reference spectra and NLS fitting) as a mixture~(red dotted line) of  79\% fibrin~(magenta line)\,\cite{Jain2014} and 21\% arachidic acid~(green line)\,\cite{Czamara2015} in both the fingerprint and CH-stretch regions.
We therefore attribute $\Comp_3$ to a mixture of fibrin protein and saturated FA, resembling arachidic acid.
This observation is in line with previous studies on fibrin deposition in plaque areas of AD mouse brains\,\cite{Paul2007}. Furthermore, the same studies have suggested that fibrin might play a central role in the initiation and progression of the inflammatory process in AD, possibly promoting neurodegeneration. Neuronal damage of AD brains correlated with fibrin deposits is supported by our observation, revealing co-deposition of arachidic acid, as a marker of oxidative damage in tissue, with fibrin.

The spectrum of $\Comp_2$ reveals an acyl chain order, given by the ratio between the intensity at 2880\,\rcm\ and at 2850\,\rcm, twice as high as $\Comp_3$, and a strong band at 1673\,\rcm, which can be attributed to the superposition of the C=O stretching vibration of the protein backbone (Amide~I band) from a high
content of $\beta$-sheet protein structure, with the C=C stretching
vibration mode in cholesterol\,\cite{Krafft2005} and in the FA chain
of \textit{E}-unsaturated FAs (\textit{trans}-configuration)\,\cite{Movasaghi2007}. Notably, the
formation of \textit{trans} double bonds~(1670\,\rcm)
results from the conversion of \textit{cis} double
bonds~(1655\,\rcm~-~not present in $\Comp_2$) during lipid peroxidation\,\cite{Muik2005}. The spectrum of $\Comp_2$ in the fingerprint region can be well described (red dotted line) by a mixture of 81\% of synthetic \Ab\,(1-42) fibrils (spectrum available only in
the fingerprint region \cite{Dong2003}), which can be associated with
the core of \Ab~plaques with predominately $\beta$-sheet protein structure~(magenta line), and 19\% cholesteryl palmitate~(green line)\,\cite{Czamara2015}. The CH-stretch region is found to be similar to the reference spectrum of cholesteryl palmitate. 
We therefore attribute $\Comp_2$ to a mixture of two chemical constituents: \Ab\,(1-42) peptide, aggregated in amyloid fibrils with $\beta$-sheet confirmation, and cholesteryl derivatives with saturated long-chain FA chains, forming lipid aggregates. The occurrence of this mixed component in the analysis indicates that the constituents are co-localized within the spatial resolution of the data, suggesting that amyloid and cholysteryl-ester lipid bodies form a tightly connected structure in the samples.
The combination of \Ab~protein and cholesteryl ester was discussed in  previous studies on cholesterol-mutant cell lines,
where it was found that \Ab~generation is up-regulated with a
selective increase of cholesteryl ester levels, regulated by
acyl-coenzyme A:cholesterol
acyltransferase~(ACAT)\cite{Puglielli2001}. The same studies  have
shown that pharmacological ACAT inhibitors sufficiently suppress
both cholesteryl ester and \Ab~biosynthesis, indicating their
potential for use in the therapy of Alzheimer's disease. Recently,
this strategy has been confirmed on an AD mice
model\,\cite{Murphy2013}. 
Furthermore, the solid lipids might be associated with lipid droplets, originated either from microglia or the lipid transporter protein apolipoprotein E (ApoE). The former is suggested by previous observations that lipid microvesicles, secreted by microglia, are shown to adhere to fibrillar plaques, from where they extract toxic
\Ab~oligomers, implying one of the possible mechanism of 
lipid deposition\,\cite{Joshi2014}. The latter is
suggested by a study showing that the ApoE-lipid complex binds
to fibrillar \Ab~plaques, where it modulates its aggregation
and clearance\,\cite{Burns2003}.

The spectrum of $\Comp_4$ shows three sharp bands characteristic of carotenoids, at 1007\,\rcm, 1154\,\rcm, and 1518\,\rcm, 
which can be assigned to the methyl rocking vibrational mode, C-C,
and C=C in-phase stretch vibrations of the polyene chain,
respectively\,\cite{Radu2016}. Furthermore, the relative intensities of the bands are comparable with those for pure $\beta$-carotene (red dotted line), confirming its attribution to $\beta$-carotene. Notably, this component is particularly enriched in cortical plaques, as will be detailed in
\Sec{sec:CompositionCortex}. Note that while average carotenoid
concentrations in control and AD brain tissue from the frontal
lobe and occipital cortex have been reported, which for
$\beta$-carotene species account for 4.5\,ng/g in Alzheimer's brains
compared to 5.5\,ng/g in normal elderly ones, there is still no
information about its spatial distribution and correlation with \Ab~plaques. Furthermore, the role of
$\beta$-carotene and other carotenoids in lipid peroxidation in free
radical-affected diseases, such as AD, cardiovascular diseases,
atherosclerosis, and cancer is controversial\,\cite{Lloret2009,Stocker2004}. For example, it was shown \textit{in vitro} that $\beta$-carotene can act as either
anti- or pro-oxidant depending on its concentration and the oxygen
partial pressure in tissue\,\cite{Burton1983}.

The spectrum of $\Comp_5$ is found to be
consistent with Fe$_3$O$_4$\,\cite{Glasscock2008} (see red dotted line). In fact, the presence of redox metal ions including iron has been observed for AD brains, making them vulnerable to oxidative damage\,\cite{RAMOS201413}.
Furthermore, it has been recently observed by electron holography
that iron nanoparticles co-localise with the cores of
\Ab~plaques\,\cite{Plascencia-Villa2016}. There is growing
evidence that a high affinity of \Ab~peptides for metal ions
results in the formation \Ab-metal complexes, and that metal ions in
turn induce the oligomerization of toxic
\Ab~species\,\cite{Zatta2009}.

\subsection{Concentration maps of chemical components in hippocampal \Ab~plaques}
\label{sec:CompositionHippo}
\begin{figure*}
\includegraphics[width=0.95\textwidth]{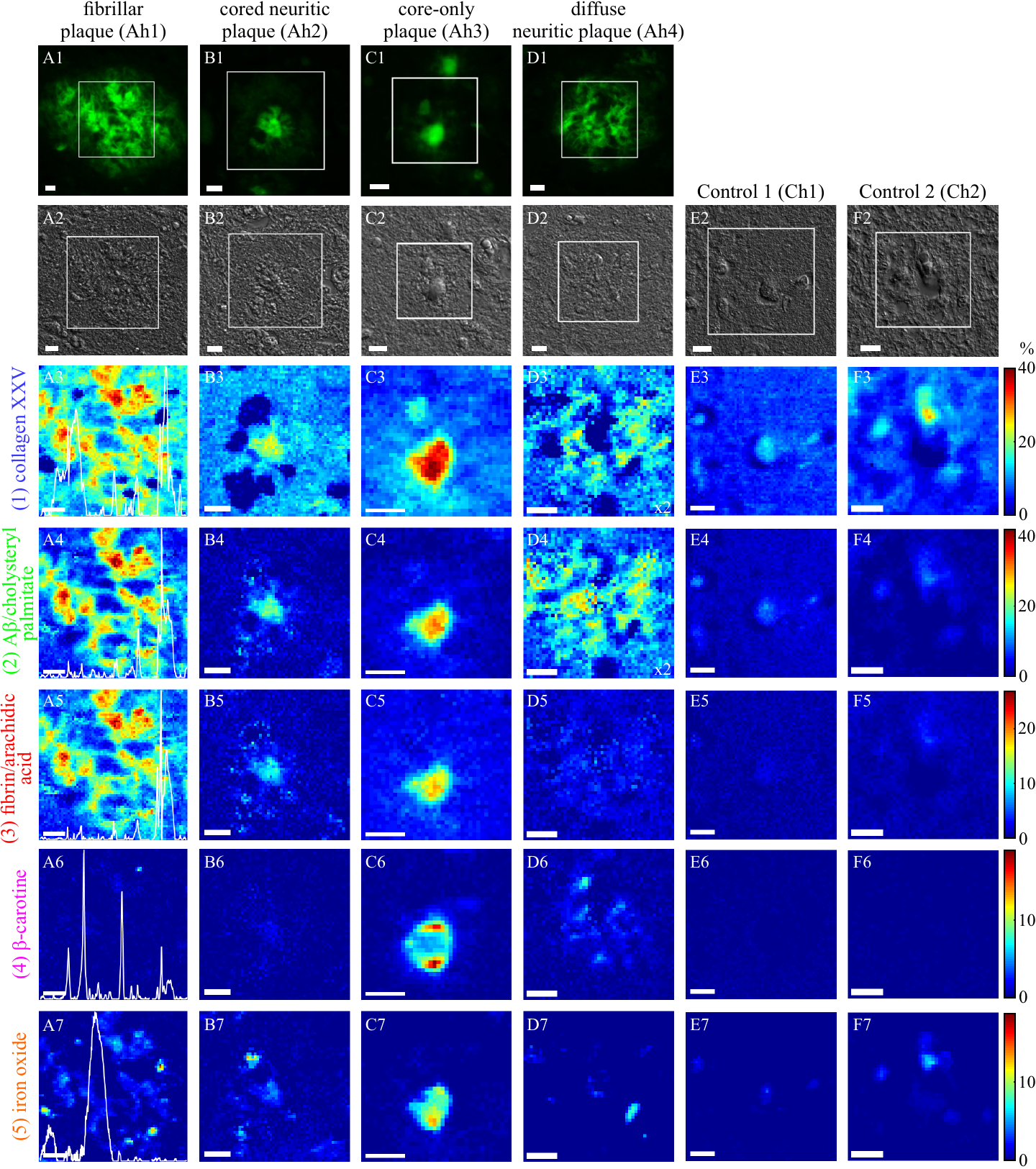}
\caption{Images of \Ab~plaques with different morphologies and of control regions. Columns A-D show fibrillar, cored~neuritic, core-only, and diffuse neuritic plaques, from Ah1, Ah2, Ah3 and Ah4 samples, respectively, and columns E-F show control regions from Ch1 and Ch2 samples. Row\,1: DIC images taken before Raman imaging. White squares indicate the regions of Raman measurements.	Row\,2: Epi-fluorescence images of the same regions as measured in DIC, taken after staining with Thioflavin-S. Rows\,3-7: spatial distributions of the concentration of $\Comp_1$..$\Comp_5$, on color scales as indicated. The concentrations in panel\,D3, D4 is scaled by a factor of two for visibility. The corresponding
component spectra are given as white lines in the first column (see also \Fig{Fig1}). Scale bars: 10\,$\mu$m.} \label{fig:RGB_Hippo_paper}
\end{figure*}

In this section, we examine and compare the spatially-resolved maps
of the concentrations of the chemical components identified in the
previous section, across AD and non-demented populations in the
hippocampal brain region. \Fig{fig:RGB_Hippo_paper} shows representative examples of \Ab~plaques with different morphologies, namely fibrillar~(\Fig{fig:RGB_Hippo_paper}A, $70\times70\,\mu$m$^2$ from Ah1), cored~neuritic (\Fig{fig:RGB_Hippo_paper}B, $60\times60\,\mu$m$^2$ from Ah2 ),
core-only~(\Fig{fig:RGB_Hippo_paper}C, $40\times40\,\mu$m$^2$ from Ah3) and diffuse neuritic (\Fig{fig:RGB_Hippo_paper}D, $50\times50\,\mu$m$^2$ from Ah4), and two control regions (\Fig{fig:RGB_Hippo_paper}E, $64\times64\,\mu$m$^2$ from Ch1, and \Fig{fig:RGB_Hippo_paper}F, $50\times50\,\mu$m$^2$ from Ch2). Row\,1 shows DIC images taken before Raman imaging, and Row\,2 shows fluorescent images of the same plaques, taken after staining with Thioflavin-S, with the white squares indicating the regions of Raman imaging. 
Rows\,3 to 7 show the concentrations of $\Comp_1$ to $\Comp_5$, respectively.
\begin{figure*}
\includegraphics[width=\linewidth]{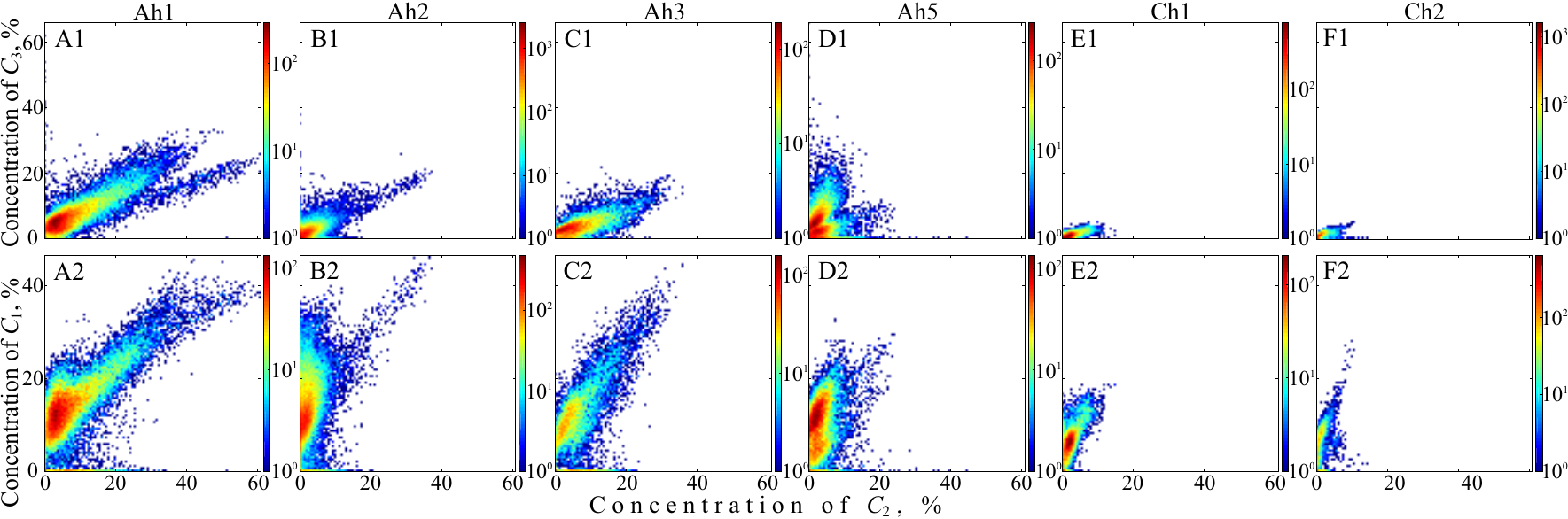}
\caption{Colocalization histograms of concentration~($\%$)
for $\Comp_2$ - $\Comp_3$ (upper row) and for $\Comp_2$ - $\Comp_1$ (lower row) for each diseased (4 AD patients) and control (2 subjects) hippocampal samples, which refer to AD and non-demented age-matched subjects, labelled according to the sample source as Ah1, Ah2, Ah3, Ah5, Ch1, and Ch2, and representing the statistics on 10, 9, 5, 5 plaques, and 6, 4 control regions, respectively, each of about 50$\times$50~$\mu$m$^2$ in size.} 
\label{fig:Fig3}
\end{figure*}
The concentration maps exhibit an accumulation of the selected components in plaque areas compared to control regions. Generally, $\Comp_1$\,(CLAC), $\Comp_2$\,(fibrillar \Ab\,(1-42) and saturated cholesteryl esters), and $\Comp_3$\,(fibrin and arachidic acid) show a similar spatial distribution.

In the fibrillar plaque, $\Comp_1$, $\Comp_2$, and $\Comp_3$ are clustered in multi-domains of various sizes, which are linked to each other, 
forming one macro-aggregate, templating the plaque. Notably, the distribution of $\Comp_5$\,(iron~oxide) is co-localized with $\Comp_1$, $\Comp_2$, and $\Comp_3$, except from small high density clusters, which seem to exclude the other components. $\Comp_4$ instead is essentially absent.

The cored~neuritic plaque, apart from being more spatially localized, shows a similar contribution of the different components, except for the collagen component, which is not observed in its
oligomeric halo, as seen from the void
areas in its concentration map~(\Fig{fig:RGB_Hippo_paper}B3), which are filled by iron\,($\Comp_5$) in the centre, and lipids\,($\Comp_3$) around.
     
The core-only plaque shows higher concentrations and a high degree of co-localization of all components. Furthermore, a significant concentration of $\Comp_4$\,($\beta$-carotene) is found surrounding the core of plaque.   

The diffuse neuritic~plaque has a similar morphology as the fibrillar plaque, but with lower concentrations of each component, except $\Comp_4$. Notably, $\Comp_2$ and $\Comp_5$ are found not to be co-localized.

In contrast to plaque areas, the two control regions have smaller concentrations of $\Comp_2$, $\Comp_3$, $\Comp_4$, and $\Comp_5$. However, $\Comp_1$ is found to have relatively high concentrations, which are rather evenly distributed, representing extracellular matrix.

\Fig{fig:Fig3} shows the colocalization histograms of the concentrations of $\Comp_2$ and $\Comp_3$ (top row) and
of $\Comp_2$ and $\Comp_1$ (bottom row) for AD and control, combining data from 10, 9, 5, 5 plaques from Ah1, Ah2, Ah3, Ah5, and 6, 4 control regions from Ch1, Ch2 individuals, respectively, each of about 50$\times$50~$\mu$m$^2$ in size. 
The colocalization  histograms for Ah4 sample is shown in the Supporting
Information Fig.\,S22. The analysis of these histograms
reveals a significant difference in concentrations of collagen, \Ab\,(1-42) fibrils, saturated cholesteryl esters, fibrin and arachidic acid between AD and control groups, indicating their potential to be used for discrimination of patients' classes. In particular, as can be seen from \Fig{fig:Fig3}, we can take a cut-off value of the distributions histograms at concentrations of 15$\%$, 5$\%$, and 15$\%$, 28$\%$ for
components~2, 3~(top row) and components~2, 1~(bottom row), respectively, which results in only histograms for AD group having concentrations higher than these values, therefore indicating that the hippocampal AD samples are differentiated from the non-AD controls on the basis of these concentration distributions.


\subsection{Raman spectra of the chemical components in cortical
\Ab~plaques}\label{Ramancortex} 
From our Raman micro-spectroscopy measurements and quantitative factorization analysis on cortical AD samples together with non-demented controls, we find four key chemical components, which are co-localized with cortical \Ab~plaques. These are: a mixture of fibrillar \Ab\,(1-42) and saturated lipids with cholesteryl derivatives~($\Comp_2$), a mixture of cholesterol and cholesteryl esters with saturated FA chains~($\Comp_3$), $\beta$-carotene~($\Comp_4$) and Fe$_3$O$_4$~($\Comp_5$) (see the Supporting
Information Figure S46). Notably, there is no signature of collagen ($\Comp_1$ in the hippocampus) observed in the cortical plaque regions. 

\subsection{Concentration maps of chemical components in cortical \Ab~plaques}

\label{sec:CompositionCortex}
\begin{figure*}
\includegraphics[width=0.95\textwidth]{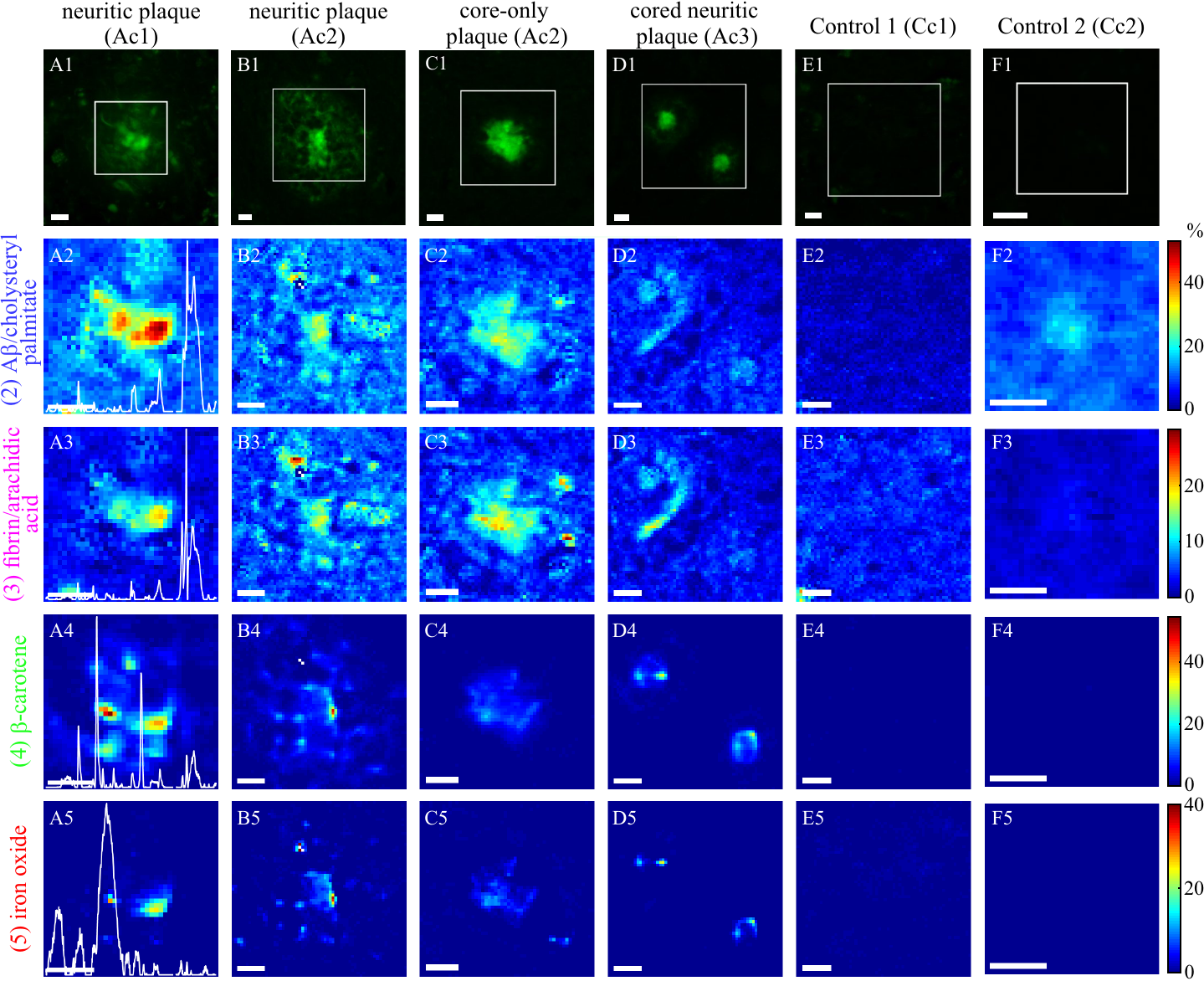}
\caption{ As \Fig{fig:RGB_Hippo_paper}, but for the cortex samples.
}\label{fig:RGB_Cortex_paper}
\end{figure*}

We compare the concentration maps of the four
components identified in \Sec{Ramancortex} over AD and
control populations in the cortex brain region.
Rows\,2-5 in \Fig{fig:RGB_Cortex_paper} show the
concentration maps, obtained from the analysis
on 31 unstained cortical \Ab~plaques (6 AD patients) together
with 9 control Raman maps (2 age-matched humans without AD). Results
are shown on examples of \Ab~plaques of three different morphologies: two neuritic~(40$\times$40$\,\mu$m$^2$ from Ac1 in column\,A, $64\times64\,\mu$m$^2$ from Ac2 in column\,B), core-only~($55\times55\,\mu$m$^2$ from Ac2 in column\,C), cored neuritic (64$\times$64$\,\mu$m$^2$ from Ac3 in column\,D), and two control regions~($64\times64\,\mu$m$^2$ from Cc1 in column\,E, and $32\times32\,\mu$m$^2$ from Cc2 in column\,F). FM of plaque and two control regions, stained with Thioflavin-S, are shown in row\,1, where white squares represent the regions investigated with Raman micro-spectroscopy. The corresponding DIC images of the same plaque and control areas are shown
in the Supporting Information Fig.\,S44. White areas in panels\,B2-B5 represent pixels which were excluded from the analysis due to too strong fluorescence.

\begin{figure*}[t!]
\includegraphics[width=\linewidth]{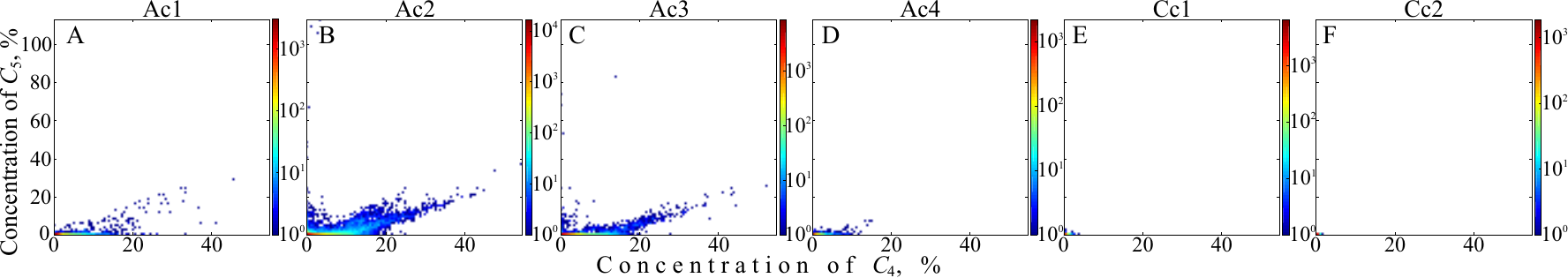}
\caption{Colocalization histograms of
concentration~($\%$) for $\beta$-carotene~($\Comp_4$) and
iron oxide~($\Comp_5$) components for each diseased (4 AD
patients) and control (2 subjects) cortex samples, which refer
to AD and non-demented age-matched subjects, labelled as
Ac1~(A), Ac2~(B), Ac3~(C),
Ac4~(D) and Cc1~(E), Cc2~(F), representing the statistics on 3, 13, 7, 2 plaques, and 4, 3 control regions, respectively, each of about 50$\times$50~$\mu$m$^2$ in size.}  
\label{fig:Fig5}
\end{figure*}

Compared to the hippocampal data, we find a similar behaviour of $\Comp_2$ and $\Comp_3$. However, the collagen component $\Comp_1$ is absent, and all plaques show large $\beta$-carotene\,($\Comp_4$) and iron oxide\,($\Comp_5$) concentrations similar to the core-only plaque in the hippocampal data. Also here, the $\beta$-carotene tends to be in the periphery of the plaque. 

We find that the concentration maps of the aggregated 
\Ab~peptides and saturated lipids with high cholesteryl
ester content\,($\Comp_2$) in the plaque areas~(\Fig{fig:RGB_Cortex_paper}(A2-D2)) correlate well with the corresponding fluorescent images of Thioflavin-S dye~(\Fig{fig:RGB_Cortex_paper}(A1-D1)).
The concentration maps of fibrin/arachidic acid~($\Comp_3$),
$\beta$-carotene~($\Comp_4$)
and iron oxide~($\Comp_5$) across
the different types of plaques show their elevated micro-scale
accumulations throughout the entire plaque areas, highest at the
fibrillar core and lowest at the \Ab~rim. In
contrast to AD plaque regions, the controls do not to contain significant amounts of $\beta$-carotene~(\Fig{fig:RGB_Cortex_paper}(E4,F4)) or iron
molecules~(\Fig{fig:RGB_Cortex_paper}(E5,F5)).
Notably, the fibrin/arachidic acid component is found not only inside the
plaque, but also outside of it, whereas $\beta$-carotene and iron
species are exclusively observed inside the plaque.

The elevated deposits of $\beta$-carotene, iron oxide, \Ab~fibrils, and saturated lipids, which are found to be co-localised in \Ab~plaques, can reflect the
presence of neuroinflammation and a high level of neurotoxicity in
the AD brain. The oxidation status of lipids is supported by the
presence of \textit{trans} double bonds in the Raman spectrum of
$\Comp_2$~(see \Sec{sec:RamanAssignment}), forming
during lipid peroxidation. Furthermore, the presence of arachidic acid~(eicosanoic acid) in $\Comp_3$ indicates the conversion of arachidonic acids to eicosanoids occurring in the process of oxidative damage. Notably, Raman bands which are
characteristic for secondary oxidation products, such as short-chain
hydrocarbons~(produced from the reaction of lipid hydroperoxide and
redox metal), are obtained when analyzing AD samples \textit{only}
in \Ab~plaque regions~(data not shown). 
Altogether, our data show that AD brains contain huge
amounts of oxidative stress markers, co-localised with
amyloid/lipids in \Ab~plaques, and thereby indicating the critical role of oxidative damage in AD pathogenesis. In terms of
\Ab~plaque aggregation, it is possible that the plaque
environment, shown to consist of abnormal micro-complexes of
fibrin/arachidic acid, saturated cholesteryl esters, $\beta$-carotene and
iron oxide, controls the degree of plaque aggregation/fragmentation and,
therefore, the neurotoxicity status of AD human brains.

Colocalization histograms of concentration are shown in
\Fig{fig:Fig5}, for the
$\beta$-carotene~($\Comp_4$) and iron oxide~($\Comp_5$) components of
each diseased (4 AD patients) and control (2 age-matched
subjects without AD) samples, using data of 3, 13, 7, 2 plaques from Ac1, Ac2, Ac3, Ac4, and 4, 3 control regions from Cc1, Cc2 individuals, each of about 50$\times$50~$\mu$m$^2$ in size.
The corresponding concentration histograms for Ac5 and Ac6 samples are
shown in the Supporting Information Fig.\,S45.
The analysis of the distribution histograms again reveals striking changes
in the concentrations of these components between AD and control
populations, indicating their potential capability to differentiate
patients' classes, similar to what discussed for the hippocampus regions (see \Sec{sec:CompositionHippo}). In particular, as can be seen from \Fig{fig:Fig5}, taking the cut-off value of the distributions histograms at concentrations of 5$\%$ and 3$\%$ for
components~4 and 5, respectively, results in only histograms for AD cohort having concentrations higher than these values, therefore again indicating that AD
population of the cerebral cortex tissue can be separated from non-demented age-matched group on the basis of these concentration distributions.

\section{Conclusion}

We have performed an extensive Raman micro-spectroscopy study coupled with quantitative
factorization analysis in AD human brain tissues. We have found
micro-scale accumulations of chemical components attributed to
cholesteryl esters with saturated long-chain FAs, \Ab~fibrils, fibrin/arachidic acid, CLAC, $\beta$-carotene and magnetite, co-localising in
\Ab~plaques of AD human brains. Notably, our study reveals
that aggregated \Ab~peptide co-exists with saturated
cholesteryl esters in one complex, representing one chemical
component. This result suggests that cholesteryl esters might play
an important role in \Ab~metabolism. Our observations support previous \textit{in vitro} studies, demonstrating that the inhibition of ACAT function, which keeps the balance between free cholesterol and cholesteryl ester levels in the brain, down-regulates \Ab~biosynthesis, therefore implying the use of ACAT inhibitors as potential therapy in the treatment of AD.
Moreover, significant accumulation and co-localisation of \textbeta-carotene and magnetite (Fe$_3$O$_4$) with lipid aggregates and \Ab~plaque cores suggests that a high level of oxidative damage is implicated in AD human brains.
Importantly, we have shown, by the means of colocalization histograms, that these chemical substances can differentiate AD brains from non-demented ones, implying their potential to be used in the diagnosis of AD.

\begin{acknowledgments}
This work was supported by the Cardiff University College of
Biomedical and Life Sciences under the International Scholarship for
PhD student. 
Part of the set-up was funded by the BBSRC grant n. BB/H006575/1. P.B. acknowledges the Royal Society for her Wolfson Research Merit Award (WM14007). We acknowledge the Oxford Brain Bank, supported by the Medical Research Council (MRC), Brains for Dementia Research (BDR) (Alzheimer Society and Alzheimer Research UK), Autistica UK and the NIHR Oxford Biomedical Research Centre.
We also acknowledge I. Pope for assistance in
the data acquisition, F. Masia for useful discussions, M. Triantafilou for assistance with samples, and M. Kukharsky for help with the staining procedure.
\end{acknowledgments}

\section*{Supporting Information}
Additional information as noted in the text.

\section*{Data Availability}
The Raman micro-spectroscopy data created during this research will be openly available from Cardiff University data archive at https://doi.org/10.17035/d.2018.0046308965 soon.

\end{document}